\documentclass[%
 reprint,
 amsmath,amssymb,
 aps,
]{revtex4-1}

\usepackage{graphicx}% Include figure files
\usepackage{dcolumn}% Align table columns on decimal point
\usepackage{bm}% bold math
\begin{document}

%\preprint{APS/123-QED}

\title{$\pi$ phase difference between Hall oscillation and SdH oscillation and non trivial Berry Phase in a topological insulator}

\author{Debarghya Mallick, Shoubhik Mandal, R. Ganesan, P. S. Anil Kumar$^{*}$}
\affiliation{Department of Physics, Indian Institute of Science, Bangalore 560012, India}

\date{\today}% It is always \today, today,
\email{anil@iisc.ac.in}
\begin{abstract}
The quantum oscillation is an important probe for the detection of a topological insulator(TI) surface states by means of electrical transport since the Shubnikov-de Haas oscillations allow to extract the Berry Phase which is the key test to detect the topological surface states. Here we have extracted the non trivial Berry Phase of 1$\%$ Sn doped strong TI $Sb_2Te_2Se$. We observed oscillation in Hall resistance as well and showed that this does not arise neither from the dominance of the SdH on Hall data nor this is the precursor of quantum Hall effect, rather this happens due to the  pinning of the Fermi Level. Also The Hall oscillation has exactly 180$^\circ$ phase difference from SdH oscillation and this phase shift is independent of the magnetic field strength. It is argued that this unusual phenomenon stems from the predominance of the intra Landau Level scattering over the inter Landau Level scattering and it depends on the strength of the scattering potential. Thus our work paves the way of understanding the physics of scattering via quantum oscillations.
\end{abstract}

\maketitle

%\tableofcontents

\section{Introduction}
Topological Insulators are the new class of materials which has brought a new arena in the community over the last decade for its exotic properties and applicability in the field of quantum computing and dissipationless electronics \cite{1,2}. From Angle Resolved Photoemission Spectroscopy (ARPES) the gapless surface states and the gapped bulk states can easily be verified \cite{3}.The magnetotransport studies using the Shubnikov-de Hass Oscillation which is the consequence of Landau Level formation is the another probe to detect the topological surface states(TSS) via the extraction of the Berry Phase\cite{4}. Berry Phase is a geometrical phase that takes the value $\pi$ when an electron encircles a "band touching" in the first Brillouin Zone and around the band touching point the dispersion is linear\cite{5}. Thus measuring the Berry Phase will be a litmus test to probe the linear band touching or the TSS in topological material\cite{4,6,7}. $Sb_2Te_2Se$ is a p-type strong topological insulator which was theoretically predicted\cite{8} and experimentally verified by APPES \cite{9}. From SdH oscillation it was confirmed to have non trivial Berry Phase \cite{14}

In this work we doped tin very minutely (around 1$\%$) to $Sb_2Te_2Se$ to create impurity band and to check its effect on the SdH oscillation since it is known that tin doping creates impurity band inside the band gap\cite{10}. We extracted the $\pi$ Berry Phase from the SdH oscillation confirming the robustness of the topological surface states against the tin doping. We observed that the Hall resistance is also oscillating on its classical value at higher magnetic field with a lesser amplitude along with the magnetoresistance oscillation (SdH oscillation). Now we qualitatively showed that the oscillation in Hall resistance is neither the precursor of quantum hall effect nor due to the admixture of SdH oscillation unlike the previous Hall oscillation study in topological insulator materials\cite{11}, rather it is the effect of the pinning of the Fermi Level \cite{12,13}

We also investigated the significance of the phase difference between the Hall oscillation and the SdH oscillation in a topological material based on the ratio of inter landau level to intra landau level scattering.

\section{Experimental methods}
%%%%%%%%%%%%%%%%%%%%%%%%
High quality single crystals of $Sb_{1.99}Sn_{0.01}Te_2Se$ were grown by the modified Bridgman method \cite{14,15}. At first, high purity Bismuth (5N), Tellurium (5N), Selenium (5N) and Tin(4N) powder were mixed and sealed in a quartz ampoule with vacuum 2 $\times$ 10$^{-5}$ mbar. Then the ampoule was kept inside the vertical tube furnace and was heated upto 850$^{\circ}$ and was kept at that temp for 3 days to homogenize and then slowly (2.5$^{\circ}$/ hr) it was ramped down upto 550$^{\circ}$ and again at that temp it was hold for 3 days to enhance the quality of the crystal. Then it was ramped down to the room temp at a faster rate. The resultant shinny crystal was cleaved easily along the c axis. X-Ray Diffraction (XRD) was done  to ensure the phase and the orientation of the single crystal. Stoichiometry was determined with the help of JEOL Electron Probe Micro Analyzer (EPMA) system. Then mechanical exfoliation was done using scotch tape technique on a $SiO_2(285 nm thick)/Si(100)$ substrate. In order to make device, we chose flat, thinner and bigger size crystals using an optical microscope. Then Hall bar designs were patterned using standard electron beam lithography technique and Cr/Au (10 nm/200 nm)contacts were given using electron beam evaporator and Samples were mounted in an oxford 2K system for the magnetotransport measurements. 

\section{Results and Discussions}The XRD data in Fig 1a displays the single crystalline phase. The (0,0,3n) orientated growth indicates that the single crystal has grown along the c-axis which is why the crystal can be cleaved easily along the c-axis. From Atomic Force Microscopy (AFM) study the thickness of the device is confirmed to be around 50 nm as shown in Fig 1b which also shows the longitudinal resistance and the Hall resistance geometry schematics along with the current probes. An  Electron Probe Micro Analyser(EPMA) was employed to record the doping percentage of Sn which yield 1$\%$ Sn replacing Sb in $Sb_2Te_2Se$.

\begin{figure*}[htb]
	\centering
	\includegraphics[width=0.80\textwidth]{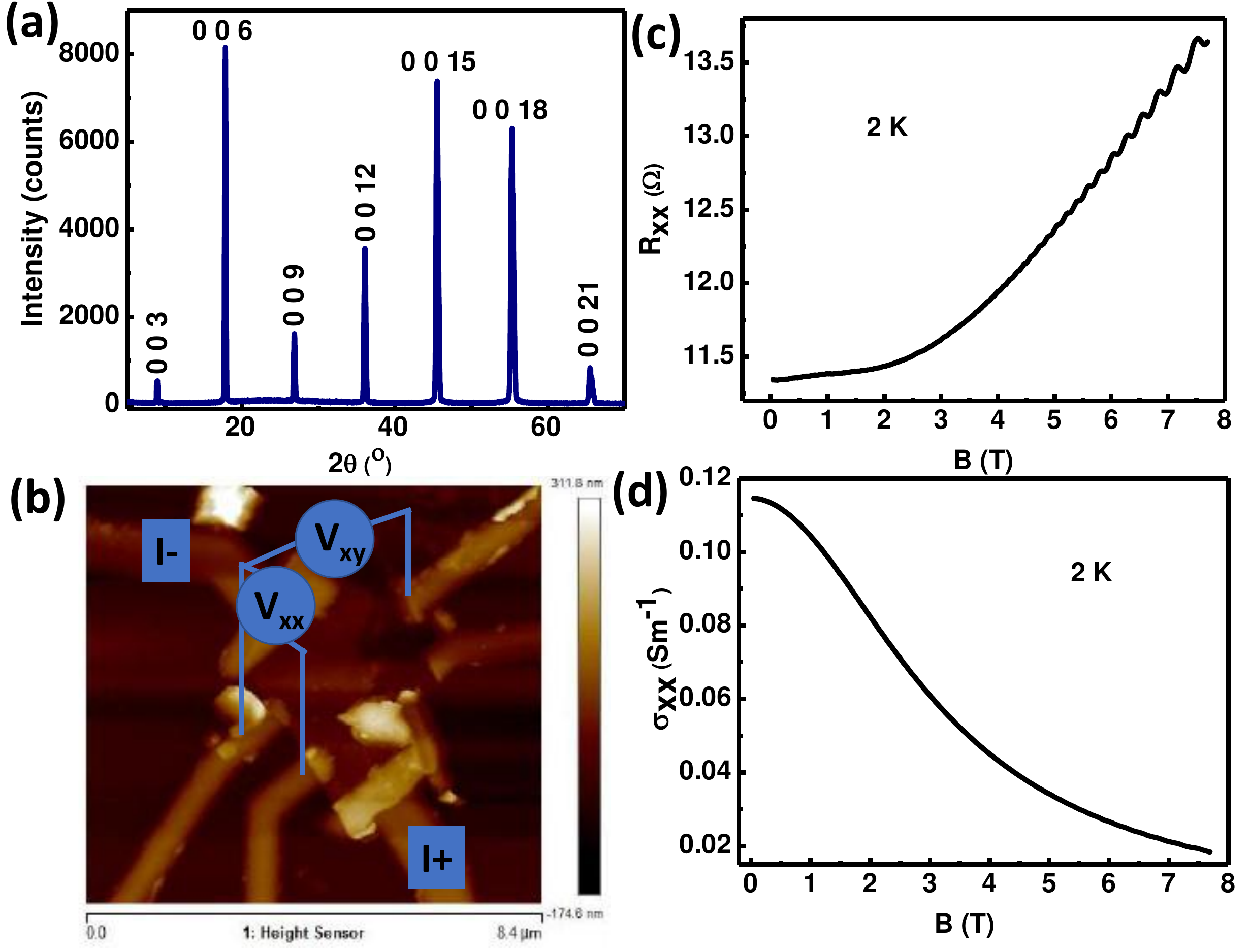}
	
	\caption{(a) XRD of a single crystal flake. (b) AFM image of the device. The longitudinal ($V_{xx}$) and the transverse (Hall,$V_{xy}$)  geometry is also shown. (c) Magnetoresistance of the device. The SdH oscillations are visible at higher field. (D) Conductivity as a function of magnetic field.}
\end{figure*}

The longitudinal resistance (magnetoresistance) ($R_{xx}$) started increasing classically with the increase of magnetic field showing no sign of Weak Antilocalisation (WAL) at the lower magnetic field (Fig 1c). This is consistent with the earlier reports of the parent compound \cite{16,17}. WAL is the consequence of $\pi$ Berry Phase. Therefore since the Berry Phase is $\pi$ in the parent compound it should have shown the WAL. Instead it just showed the normal classical behavior.
To exhibit the quantum interference effect (WAL or WL), the phase coherence time (the time scale which determines how long an electron can move without losing its' phase) has to be more than the scattering time specially by the short range scattering potentials \cite{18,19}. Now among the various types of scatterer, vacancies are the one of the prime short range scatterers \cite{18}. It is well known that Te based TI materials contains sizable amount of vacancy defects\cite{20}. Here, since the flake under study is relatively thicker (50 nm) the bulk will contribute appreciably in parallel conduction due to the large content of Te vacancies in the bulk and thereby suppressing the quantum interference effect (WAL) despite having $\pi$ Berry Phase.

\renewcommand{\topfraction}{1}
\renewcommand{\bottomfraction}{1}
\renewcommand{\textfraction}{0.09}

The MR at 2K shows the SdH oscillations and it starts appearing at $\sim$ 4.5 T indicating that the formation of Landau Levels starts at that magnetic field regime. To reach the extreme quantum limit that is in the quantum hall regime three conditions are to be satisfied \cite{21}: (1) cyclotron frequency($\omega_c$) must be much higher than the scattering rate($\tau$) that is $\omega_c\tau>>1$. So, the broadening of Landau level due to the scattering should be much less than the spacing of the landau levels. (2) the spacing between two consecutive LLs has to much higher than the thermal broadening of each LL. (3) The Fermi level should be higher than at least few LL that is $E_f>\hbar \omega_c$. In other word this condition sets the maximum value of the magnetic field beyond which there will be no more oscillations since at that critical field all the electrons will occupy the lowest LL. 
The longitudinal resistance $(R_{xx})$ has been converted into Longitudinal conductivity $(\sigma_{xx})$ with the help of conductivity tensor
\begin{math}
\sigma_{xx}=\frac{\rho_{xx}}{{\rho_{xx}}^2+{\rho_{yy}}^2}
\end{math} (Fig 1d). The reason we dealt with the conductivity instead of resistivity is that the result might have been misleading if we take resistivity while analyzing the LL fan diagram (will be discussed later in Fig 3c) depending on the absolute value of $\sigma_{xx}$ and $\sigma_{xy}$ \cite{22}. As a consequence one can see that the amplitude of the oscillations in the conductivity is reduced compared to that in the resistance data. 

Fig 2a shows the SdH oscillations after subtracting the background from the magnetoconductivity curve and the conductivity oscillates periodically in 1/B. It is evident that the amplitude is getting lower with the increase in temperature. This is not surprising because increasing the temperature will increase the thermal broadening which in turn will reduce the SdH oscillation amplitude. Now to fit the SdH oscillations and the temperature dependence of its amplitude, we have used the Liftshitz-Kosevich (LK) equation \cite{23,24}: 
\begin{equation}
\sigma_{xx}\propto\frac{\lambda}{\sinh\lambda} \mathrm{e}^{- \lambda_{D}} \cos2\pi\{\frac{F}{B}+\gamma-\delta\}
\end{equation}
where,
\quad{$\lambda=\frac{2\pi^2K_BTm^*}{\hbar eB}$} and $\lambda_D=\frac{2\pi^2K_BT_Dm^*}{\hbar eB}$\\

Here F is the Fast Fourier Transform (FFT) value of the oscillations plotted in the reciprocal magnetic field (1/B). $2\pi\gamma$ is the Berry Phase and delta is the dimensional parameter. $\delta$ takes value $\pm$1/8 for 3D and for 2D it is 0 \cite{24}. In our case the material being a strong TI possesses 2D topological surface states which allows us to take value of $\delta$ = 0. $T_D$ is the Dingle temperature (we discussed this in the next paragraph). Thus it is obvious from the LK equation that the first two terms are there to take care of the damping part of the oscillation whereas the last part (the cosine function) takes care of the oscillatory nature.

From the temperature dependence of the oscillation amplitude we can calculate the effective mass or the cyclotron mass of the Dirac electron whereas the magnetic field dependence will yield the Dingle time and LL broadening \cite{25}. Thus we fit the temp dependence at a fixed magnetic field (7.315 T) with $\sigma_{xx} \propto \lambda/ sinh \lambda$ which gives the effective mass as 0.11 $m_e$ where $m_e$ is the rest mass of the electron(Fig 2b). Then we fit the mag field dependence of the amplitude at a fixed temp (2 K) with $\sigma \propto \mathrm{e}^{-\lambda_D}$ and extract the Dingle temperature ($T_D$) to be around 14 K and Dingle time ($\tau_D$) to be $9\times10^{-14}$ sec (Fig 2c). These values are consistent with the previous reports in TI material \cite{11,25,26}. The Dingle time basically is the mean time interval between two successive collisions and LL broadening can be calculated from the Dingle temperature as $\Gamma=\pi K_B T_D=4.76meV$ \cite{27}.

 \begin{figure*}[!htb]
	\centering
	\includegraphics[width=1.00\textwidth]{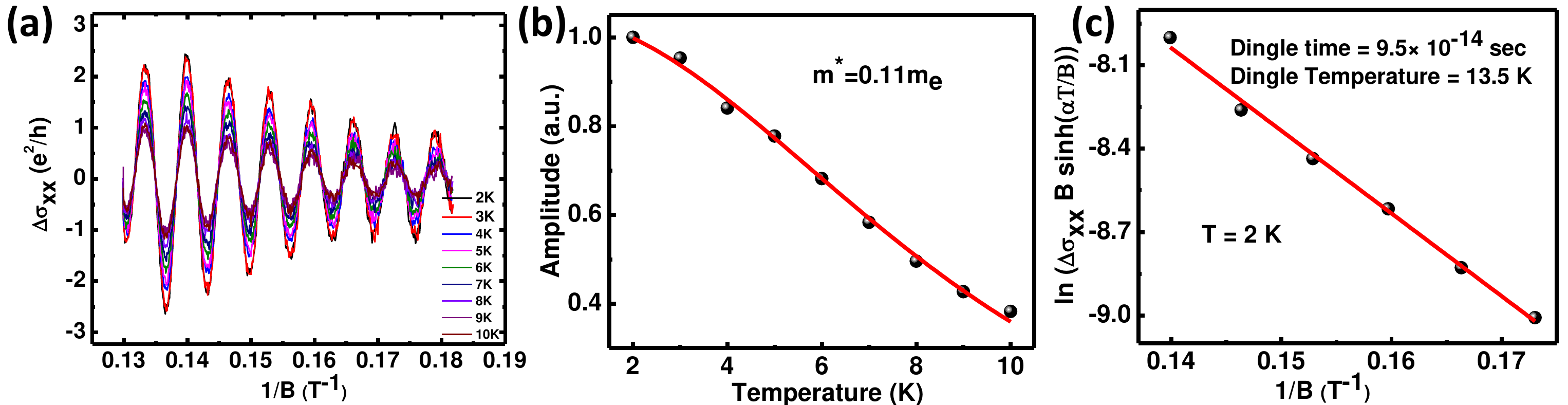}	
	\caption{(a) SdH oscillations in the reciprocal magnetic field at different temperatures. (b) Temperature dependence of the SdH oscillation amplitude at a particular magnetic field. Fitting with temperature part of the Liftshitz-Kosevich formula yields the effective mass of the electron. (c) Dingle plot with linear fitting which yields the Dingle time and the Dingle temperature.}
\end{figure*}

33Now from the oscillatory part of the LK formula that is the cosine part, we extract the Berry phase to understand if the electron is behaving as Dirac Fermion. Here $2\pi \gamma$ is the Berry Phase.  We extract the Berry Phase by two ways to check the self consistency. First we fitted the SdH oscillation with the LK formula extracting the Dingle Temperature to be around 17K which is close to the value obtained from the mag field dependence plot (14K) and the Berry Phase ($2\pi \gamma$) to be 0.41$\times$2$\pi$=0.81$\pi$. Secondly we plotted the Landau Fan Diagram which is the LL index vs the inverse of magnetic field (1/B) plot (Fig 3b). Here while indexing the LLs, we followed the convention of assigning the integer value to the minima of the oscillations \cite{16}. The intercept of the extrapolated straight line to the LL index axis is of prime interest since this manifests the Berry Phase. So with the extrapolation basically we seek to see what happens to the Fermi Level position when the magnetic field tends to infinity which means all the electron are residing in the lowest LL (LLL). Now the specialty of the Dirac material is having the zeroth LL which is equally shared by both the electrons and the holes \cite{28}. Thus the physical picture of the intercept is basically to find the existence of the LLL. Now Berry Phase generally for non-Dirac material takes the value zero, but if the band structure of the material is having Dirac cone then the Berry Phase is $\pi$ \cite{5,7}. So the value of the Berry phase can demonstrate the Dirac cone or the topological nature of the electron inside a topological insulator material. Now it can be readily shown from the argument part of the cosine function that if the intercept value in the Fan Diagram is 0.5 that means Berry Phase is $\pi$ \cite{22} or it holds the topological nature and if the value is zero that indicates that the electrons are no more Dirac Fermion. It is clear from Fig 3c that in our case the intercept is 0.4 $\pm$ 0.05 which is close to the value derived from the LK fitting of the SdH oscillation. Now it is worthy to mention that the intercept value is not exactly 0.5 and it is very rare in the literature reporting exactly 0.5 \cite{11,24,26}. The reason behind this deviation is the Zeeman coupling which we were not taking into account so far. If the Zeeman coupling is strong enough then the there will be an extra term added to the Dirac Hamiltonian \cite{29}. Thus the linear band dispersion at lower energy is somewhat now compromised and that in turn will reflect in the value of Berry Phase. We can extract the Fermi wave vector using the Onsager relation, F=$\frac{\hbar}{2e} K_F^2$ where F is the FFT value and $K_f$ is the Fermi vector. $K_f$ is estimated to be 
0.0678$\AA^{-1}$. Fermi velocity($v_F$) and mean free path ($l_e$) also we extracted from the oscillation as $7.15\times10^5$ m/sec and 67 nm respectively using the relations $v_F=\frac{\hbar}{m*}K_F$ and $l_e=v_F\tau_D$.\cite{11,26} 
 \begin{figure*}[!htb]
	\centering
	\includegraphics[width=0.75\textwidth]{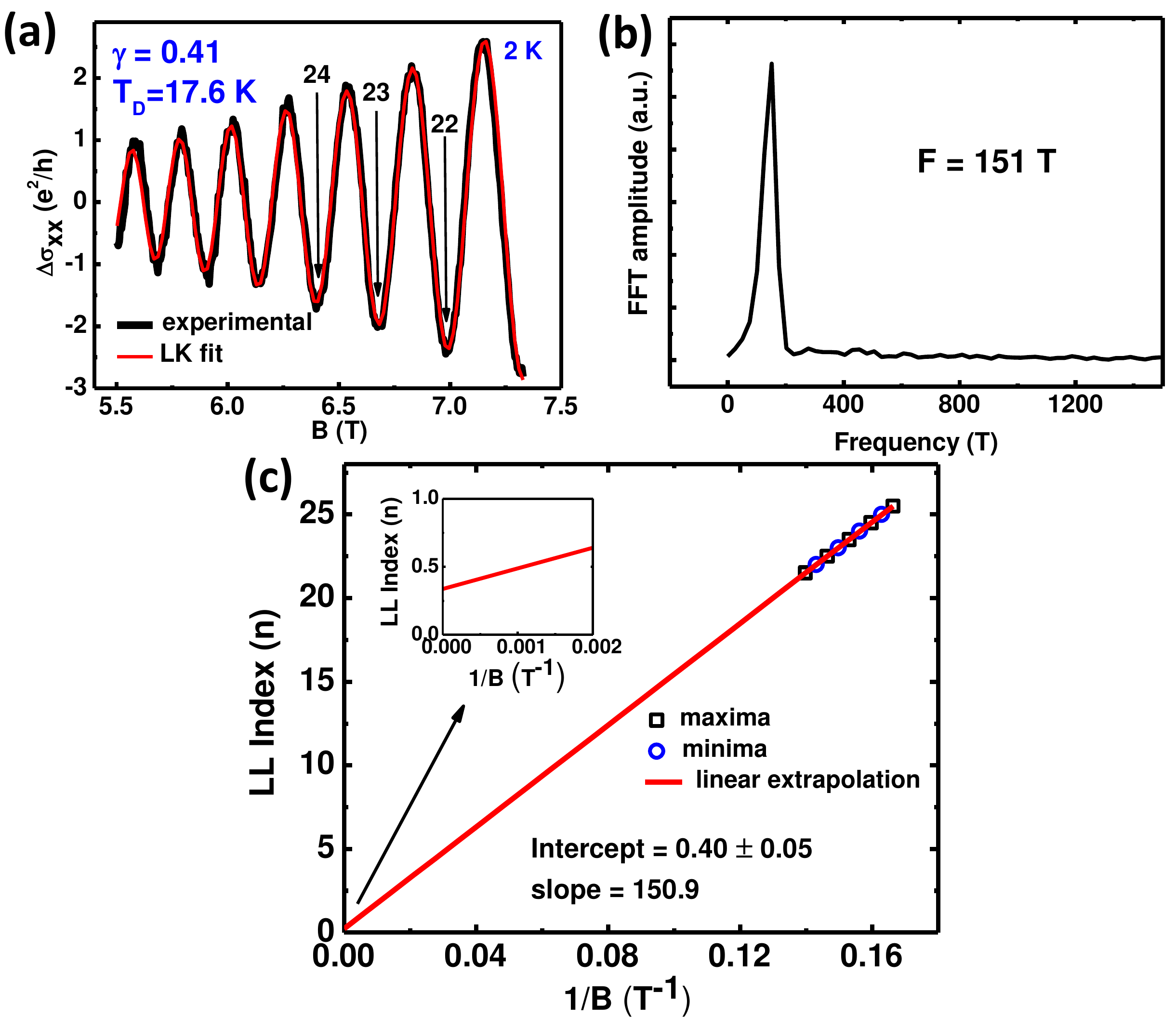}	
	\caption{(a) Liftshitz-Kosevich fit to the SdH oscillation at 2 K. Fit yields the Berry Phase and the Dingle Temperature. The integer value has been assigned to the minima in the SdH oscillation. These values are the Landau Level indices. (b) The Fast Fourier Transform  corresponding  to the oscillation. (c) Landau Level Fan Diagram which is the plot of the LL indices with the reciprocal magnetic fields. maxima are assigned the half integer value whereas the mimina are assigned the integer value. The intercept of the extrapolated straight line provides the Berry Phase and the slope should match with the FFT value. Inset is showing the zoomed view of the intercept.)}
\end{figure*}
\begin{figure*}[!htb]
	\centering
	\includegraphics[width=0.80\textwidth]{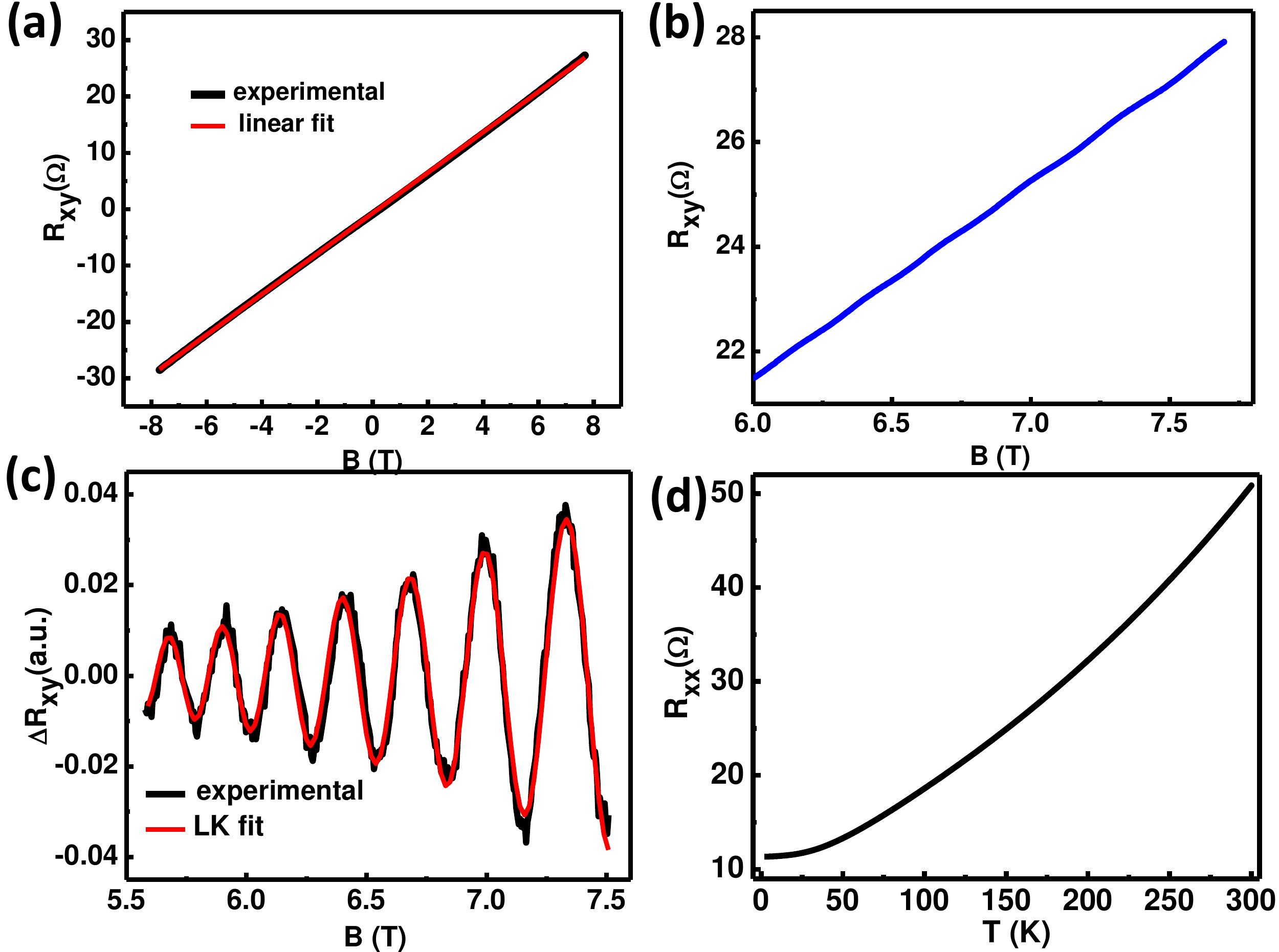}	
	\caption{(a) Linear Hall effect indicating hole as the carrier. (b) Zoomed in image of the same Hall data at the higher magnetic field to show the mild oscillations. (c)After background subtraction it gives prominent oscillations. Same LK fit is performed here also, showing the same Berry Phase as extracted from the SdH oscillations. (D) Resistance vs Temperature plot}
\end{figure*}

The linear Hall data (Fig 4a) indicates the single type of carrier and the positive slope tells that the carrier is hole. We can calculate the Hall mobility and the carrier concentration to be 4400 $cm^2/Vs$ and $1\times 10^{12}/cm^2$. Though SdH mobility and carrier concentrations is estimated to be 1500 $cm^2/Vs$ and $3\times 10^{14}/cm^2$ respectively. Thus the SdH mobility is almost three times less than the same derived from the Hall effect. The reason is that the SdH mobility takes care of the scattering at all the angles whereas scattering only at large angles that is only the large energy transfer scattering contributes to the Hall mobility and thereby making the Hall mobility greater than the SdH mobility \cite{4}. The discrepancy in the carrier concentration value between the Hall and the SdH stems from the fact that the Hall effect takes care of the multi band effect whereas the SdH carrier concentration takes only the single band contribution into account.\cite{21}  

Fig 4b is the zoomed image from 6 T to 7.7 T of Fig 4a where we observe that there is oscillation of the Hall resistance on top the classical Hall data and the phase shift between the Hall oscillation and the MR oscillation (SdH oscillation) is exactly $180^\circ$ since the the maxima of $\Delta R_{xx}$ is exactly merging with the minima of $\Delta R_{xy}$ (Fig 5) . Now the Hall oscillation in the context of TI is rare. Cao et al found the Hall effect to be quantized in $Bi_2Se_3$ \cite{31} and Busch et al \cite{11} got oscillations in Hall data and they also attribute it to be the precursor of the QHE. But neither of them commented about the phase shift between the SdH and the Hall oscillation.  

The reason behind the origin of the Hall resistance oscillation is somewhat controversial. In topological insulator literatures, the reports on Hall oscillation are very less and they all attributed it to the precursor of quantum Hall effect \cite{30,31}. In past in 2DEGs and narrow band gap semiconductors, the quantum oscillations in Hall effect have been investigated both theoretically and experimentally. Now generally three prime reasons are attributed for Hall oscillations. We will now discuss one by one to check which one fits the best in our case. (1) \textbf{The precursor of Quantum Hall Effect :} As we have seen that this is the mostly attributed in case of TI. But the main requirement for entering into QHE regime is $\omega_c\tau>>1$. Now taking the effective mass to be 0.11$m_e$ and the critical field at which the oscillation starts to be 4.5 T we got the value for $\omega_c\tau=0.5$ which is far from the requirement of QHE regime or even precursor to that. Another point which is worth to mention that if the Hall oscillations are really appearing as the precursor to the QHE then the phase difference between the Hall oscillation and the SdH oscillation should be around $90^\circ$ \cite{32} because the inflection in Hall resistance coincides with the SdH maxima whereas we are getting excat $180^\circ$ phase difference. These led us into the conclusion that the observed Hall oscillation is not the precursor of QHE. (2) If there is strong SdH oscillation then \textbf{admixture of $R_{xx}$ and $R_{xy}$} sometimes may give rise to the oscillations in Hall data. This may appear when the exfoliated flake is irregular in shape and the Hall contact probes are misaligned. Thus to eliminate the MR contribution ($R_{xx}$) from the Hall data ($R_{xy}$), we collected Hall data in both the magnetic field directions (+z and -z direction) and then we anti-symmetrise them as $R_{xy}=(R_{xy,raw}(+B)) - R_{xy,raw}(-B))/2$ \cite{29}. Moreover if at all the MR oscillations contribute to the Hall data then there should not be any phase shift between them as both the maxima and minima will happen at the same magnetic field. Therefore it is obvious that admixture of MR and Hall is not the case here. (3) Thus we explain the observed Hall oscillation using the model of \textbf{pinning of Fermi level}\cite{12,33}. If impurity bands are formed in the band gap then due to their high density of states the Fermi Level is pinned in those levels. These bands act like a "charge reservoir". Now at sufficient strong magnetic field oscillatory transfer of charge carrier occurs from this impurity band to the conduction band giving rise to the oscillation in Hall resistance. Yep and Bekkar showed that the necessary condition for this phenomena to happen is having high density of states of the impurity band \cite{12,21}. Here in our case the tin(Sn) doping forms this high density of states impurity band. Kahn and Frederikse \cite{34} showed that for $E_F > \hbar \omega_c > K_BT$ (For our sample this is being satisfied since $E_F$ = 158meV, $\hbar \omega_c$ = 13meV and $K_BT$ = 0.2 meV), the carrier concentration oscillation can be expressed with the same functional form as in LK formula but without the LL broadening part used for SdH oscillation:
\begin{equation}
\frac{\triangle n}{n_0}\propto\frac{\lambda}{\sinh\lambda} \cos2\pi\{\frac{F}{B}+\frac{1}{8}\}
\end{equation}

It is important to note that the Berry Phase was not included in that early time and the equation was derived for 3D case which gives the value 1/8 as discussed before (Equation 1). Now if the Landau Level broadening part (that is the Dingle temperature part) and Berry Phase is taken into account then it exactly takes the form of LK formula (equation 1) for 2D Topological surface states. Thus we fitted the Hall oscillation data with the same LK formula and found out the Berry Phase is same as derived from the SdH oscillation. Resistance vs temperature curve (Fig 4d) indicates the metallicity of the sample.

After obtaining same berry phase from the Hall oscillation and the SdH oscillation it is pertinent to ask what the phase difference is between these two oscillations and what is its physical significance.  Though Hall oscillations are reported in few topological insulator literatures,  the phase difference study between them is completely missing. Here we discuss the significance of the phase difference. Adams and Holstein discussed about this phase difference in as early as 1954 \cite{35}. They added higher order scattering terms in the expression of $\sigma_{xx}$ and showed the phase shift between $\Delta\sigma_{xx}$ and $\Delta\sigma_{xy}$ to be around $45^\circ$. But in our Sn doped topological insulator sample the phase difference is $180^\circ$ because the minima of SdH oscillations is superposing exactly with the maxima of the Hall oscillations. Previously the semimettalic HgTe system has shown $180^\circ$ phase difference \cite{36}. Mani et al prepared several narrow band gap semiconductor samples and showed that the phase shift can vary from sample to sample depending on the range of the scattering potential  \cite{32}.

\begin{figure}[!htb]
	\centering
	\includegraphics[width=0.48\textwidth]{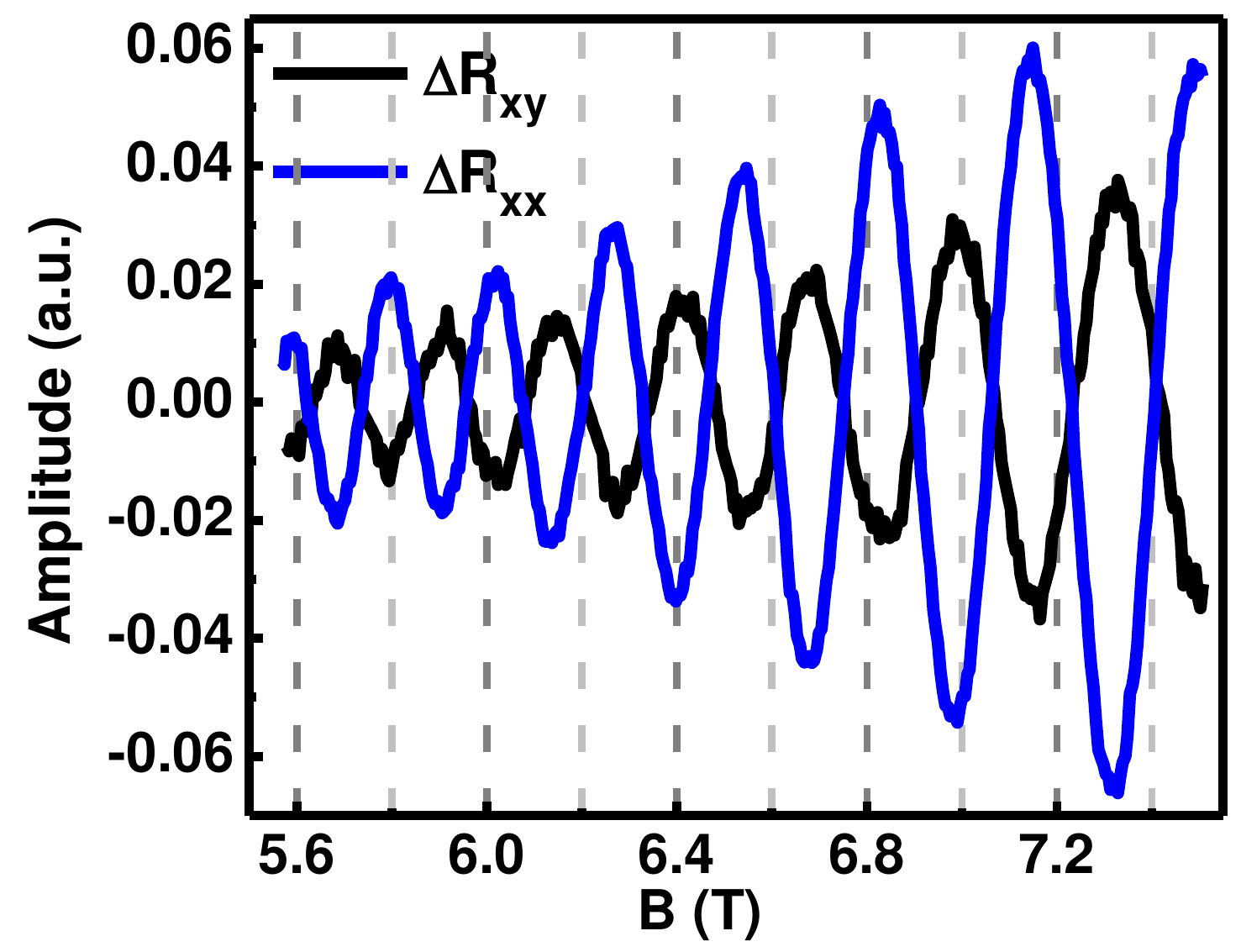}
	\caption{ Superposition of 2 K SdH oscillations (Longitudinal) and Hall oscillations (Transverse) shows the unusual $\pi$ phase shift between them and that is independent of the magnetic field. The vertical dotted lines are for guide to eye.}
\end{figure}

Thus depending on the long range or short range of the scattering potential (long range scattering potential is defined as the potential range should be $>>$ magnetic length) there can be two types of scattering: inter LL scattering and intra LL scattering. Since with the magnetic field the LL gap changes, the inter LL scattering depends on the magnetic field whereas the intra LL scattering is independent of the magnetic field. Thus in our case since the phase difference is independent of the magnetic fields we can say the scattering mechanism dominating in our sample is the intra LL one. The energy corresponding to any LL for topological material is given by \cite{28}\\
$E_n$=$v_F$ sgn(n)$\sqrt{2e \hbar |B||n|}$, \\ where $v_F$ is the Fermi velocity, sgn(n) is the sign function, e,h,B are the electronic charge, Planck's constant and the magnetic field and n is the corresponding LL index. Thus the energy gap between two consecutive LLs can be written as \\
$\Delta E_{n_1,n_2}=v_F \, {2e \hbar} \, [\sqrt{n_1B_1}- \sqrt{n_2B_2}]$ \\where $n_1,n_2$ are the two consecutive LL index at the corresponding mag field $B_1$ and $B_2$. Thus in this case, assigning $n_1$, $n_2$ as 23rd and 22nd LL, the value of $\Delta E_{23,22}$ is calculated to be 0.074 meV whereas the magnetic field independent LL broadening is $\Gamma=\pi K_B T_D=4.76 meV.$ Since $\Gamma>\Delta E_{LL}$ the intra LL scattering is much stronger than inter LL scattering. Thus the magnetic field independent $180^\circ$ phase difference in the Sn doped strong topological insulator sample is attributed to the stronger intra LL scattering and the strength of the scattering potential in our sample. 

\section{Conclusion}
In summary we have grown high quality 1$\%$ Sn doped $Sb_2Te_2Se$ single crystal which has been characterized by XRD and the stoichiometry was estimated using EPMA. Magnetotransport investigation showed SdH oscillation and Hall resistance oscillation. Non-trivial $\pi$ Berry Phase extracted from the SdH oscillation via Landau Fan Diagram and from the fitting of the Liftshitz-Kosevich formula proves the topological nature of the material. We have argued that the oscillation in Hall resistance is not arising from the SdH effect, rather the pinning of Fermi Level due to Sn doping is responsible for it. The superposition of both the type of oscillations shows a striking feature of $\pi$ phase difference between them which is not changing with the magnetic field. we discussed this as the effect of predominance of intra LL scattering and the particular strength of the scattering potential in this specific sample. Nevertheless, the paper does not answer two specific questions : (1) What exactly are the scatterers present in our Sn doped $Sb_2Te_2Se$ sample which give rise to such strong intra Landau Level scattering ? and (2) How the ratio of the intra LL and inter LL scattering are mathematically related to the phase shift of the SdH oscillation and the Hall oscillation ? Therefore this work necessitates the need of detailed theoretical investigation. We also calculated different transport parameters such as Fermi Wave vector, Fermi velocity, mean free path, mean scattering time (Dingle time). We extracted the mobility and the carrier concentration from both the Hall measurement and the SdH oscillation and compared them. Thus we believe that our work paves the way for more detailed research regarding both the types of oscillation and the phase difference between them and their interlink with topological surface states, if any.

\vspace{4mm}

\section*{Acknowledgements}

%\hspace{-7mm}\textbf{Acknowledgement}\\%[1.0ex]
D.M. and S.M. thank MHRD, Govt. of India. P.S.A.K. acknowledges Nanomission, DST, India for financial support. The authors thank NNFC and MNCF, Centre for Nano Science and Engineering and Advanced Facility for Microscopy and Microanalysis (AFMM) at the Indian Institute of Science Bangalore for fabrication and characterization.

\end{document}